# A NOVEL APPROACH FOR SECURITY ISSUES IN VOIP NETWORKS IN VIRTUALIZATION WITH IVR

Kinjal Shah[1], Satya Prakash Ghrera[1] and Alok Thaker[2]

[1]Department of Computer Science Engineering & Information Technology, Jaypee University of Information Technology, Waknaghat, Distt Solan, (H.P), India
`kinjal.93@gmail.com, spghrera@rediffmail.com`
[2]Inferno Solutions, Vadodara, (Gujarat), India
`alok.akki@gmail.com`

## ABSTRACT

*VoIP (Voice over Internet Protocol) is a growing technology during last decade. It provides the audio, video streaming facility on successful implementation in the network. However, it provides the text transport facility over the network. Due to implementation of it the cost effective solution, it can be developed for the intercommunication among the employees of a prestigious organization. The proposed idea has been implemented on the audio streaming area of the VoIP technology. In the audio streaming, the security vulnerabilities are possible on the VoIP server during communication between two parties. In the proposed model, first the VoIP system has been implemented with IVR (Interactive Voice Response) as a case study and with the implementation of the security parameters provided to the asterisk server which works as a VoIP service provider. The asterisk server has been configured with different security parameters like VPN server, Firewall iptable rules, Intrusion Detection and Intrusion Prevention System. Every parameter will be monitored by the system administrator of the VoIP server along with the MySQL database. The system admin will get every update related to the attacks on the server through Mail server attached to the asterisk server. The main beauty of the proposed system is VoIP server alone is configured as a VoIP server, IVR provider, Mail Server with IDS and IPS, VPN server, connection with database server in a single asterisk server inside virtualization environment. The VoIP system is implemented for a Local Area Network inside the university system.*

## 1. INTRODUCTION

VoIP (Voice over Internet Protocol) is a booming technology since last few years and has gained admiration in the professional and educational industries. The VoIP technology is gaining this popularity due to its open source availability to anybody from the source of the internet. It has proved itself one of the best alternate to the Public Service Telephone Network (PSTN) line telephone instruments. On implementation of this technology provides the common wiring set up for the computers as well as phone lines for the communication [6]. The technology provides a good alternate intercom facility using computers rather than hard core telephone instruments. The hard core telephones can be replaced by X-Lite kind of the soft phones or even with the IP phones which are good examples of the soft phones those can be installed on any platform. This technology interacts with both the local and remote VoIP phones using internet as well as intranet for an organization. Even the VoIP phones can be also connected with PSTN telephones for communication as well as for IVR implementation on hard core telephone lines for those kinds of organizations which only want telephone like certain units of telecom industries. The main protocols for the implementation of this technology are SIP and H.323 [10]. However, the SIP server is used for the used for configuring the VoIP server. If two different SIP servers want to register with each other from two different buildings, IAX protocol is used for this kind of the connection. The other protocols which are used for the implementation of this kind of service are Real Time Protocol (RTP), STUN and Cisco VoIP [10] [13]. In the proposed system architecture the system is implemented inside a LAN using VMware server's bridge networking facility. The VoIP provides such an immense flexibility for inter user communication inside the organization among the employees; however, the security vulnerabilities are still possible on VoIP networks. The attacker can execute the various kinds of attacks on the VoIP server to disturb its service as well as the service of IVR. These threats come under following classifications namely Confidentiality, Availability, Authenticity, Larceny, SPIT (Voice Spam). The confidentiality threats classify in to Call Eavesdropping, Call recording, and voicemail tampering. The availability threats fall in to Denial of Service (DoS) floods, Buffer Overflow attacks, Worms and Viruses. The authenticity attacks include the registration hijacking, caller ID spoofing, sound insertion. The Larceny threats consider service theft like toll fraud and data theft like masquerading data as voice and invalid data network. Finally SPIT attacks categorize unsolicited calling, voice mailbox stuffing and voice phishing [6] [7] [9]. These kind of attacks must be prevented those can disrupt the services of the VoIP networks. In the proposed paper firstly the VoIP network has been implemented with IVR facility and then system is configured with certain security parameters like Virtual Private Network, firewall, Intrusion Detection and Prevention System to protect against some of the serious attacks like Denial of Service attack, port scanning, registration hijacking and the possible attack to the database server in very much well-organized manner. The main piece of cake in the proposed architecture is the asterisk server acts as VoIP server, Mail server and VPN server along with the connection with the firewall, Intrusion Detection & Prevention System and MySQL database server. In this paper section 2 explains the related work, section 3 focuses on the proposed work includes VoIP System Implementation, IVR configuration, Mail Server accomplishment, Configuration of security parameters, section 4 enlightens the pros and cons of the proposed system section 5 consisting of practical snap shots and section 6 finally reaches to the conclusion.

## 2. RELATED WORK

In [1] the authors have examined the anonymity for QoS sensitive applications on mix networks using peer to peer VoIP service as a sample application. A peer-to-peer VoIP network typically consists of a core proxy network and a set of clients that connect to the edge of this proxy network. This network allows a client to dynamically connect to any proxy in the network and to place voice calls to other clients on the network. In [4] the authors have concentrated on the performance of VoIP network under the DoS attack by categorizing the network into SIP dependent performance matrix and SIP independent matrix. SIP dependent matrix includes parameters like Call Completion Ratio (CCR), Call Establishment Latency (CEL), Call Rejection Ratio (CRR) and number of retransmitted packets (NRR). SIP independent matrix includes parameters like CPU usage, CPU interrupts rate and Interrupt handling time. In [8] the authors have focused mainly on SIP based secure communication based on Secure Real Time Protocol (SRTP) which provides security services for Real Time Protocol (RTP) media and is signaled by use of secure RTP transport in Session Description Protocol (SDP). The authors have explained how RFC4568 defmes a SDP cryptographic attribute for unicast media streams for a VoIP network. VoIP uses the two main protocols: route setup protocol (RSP) for call setup and termination, and real-time transport protocol (RTP) for media delivery. The authors have focused on VoIP Route Set up Protocol in peer to peer VoIP networks and flow analysis attack exploit the shortest path nature of the voice flows to identify pairs of callers and receivers on the VoIP network. In [2] [3] [5] [10] [11] [13] the authors have concentrated mainly on the various security vulnerabilities on VoIP network like IP network security vulnerability, Denial of Service (DoS) attack, Service steal threat, Interception and tempering with VoIP packets, Middleman attack, Web spoofing, unauthorized access, masquerading, call hijacking. The solution provided to avoid these kinds of attacks is to follow the security strategies like formulating relevant laws and regulations, establishing separate firewall, packet encryption and authentication, ensuring the integrity and confidentiality of data packets [11]. In [3] [6] [15] the authors have also payed attention on the H.323 protocol and its system architecture. The main components of the SIP based systems are User agents (UA) and servers. User Agents (UAs) are combinations of User Agent Clients (UAC) and User Agent Servers (UAS). A UAC is responsible for initiating a call by sending a URL addressed INVITE to the intended recipient. A UAS receives requests and sends back responses. The servers can be classified in to proxy servers, redirect servers, location servers and registrar server [5] [12] [15]. In [5] the authors have focused on the insufficiency of SIP security mechanisms which are certification attack, DoS attack and spam attack. In [6] the authors have proposed the solution for defense against various mentioned attacks like separation of VoIP and Data traffics, Configuration authentication, signaling authentication and media encryption. In [9] authors have focused upon the security threats and assessment on the VoIP network. The attacking tool for attacking on the VoIP network is developed with the help of XML files. In [11] the authors have concentrated on the various VoIP attacks and its preventing policies according to NIST report. The authors have proposed three design patterns to secure the VoIP network those include secure traversal of firewalls for VoIP, detecting and mitigating DDoS attacks targeting VoIP, securing VoIP against eavesdropping. The firewall strategy provides solution for maintaining separate Global Directory Index (GDI) for online clients. The detecting and mitigating DDoS attack strategy provides solution that the communication between Media Gateways (MG) and Media Gateway Controller (MGC) must be in the form of transaction so every transaction will have unique ID. The system must be configured with Intrusion Detection System (IDS) and Intrusion Prevention (IPS) system. In the system on completion of communication the BYE message should be sent by the party that wants to terminate the connection. The eavesdropping strategy focuses on implementation of DES encryption algorithm in CBC mode. In [14] the authors have proposed various VoIP communication scenarios those include hosted services and trunking VoIP service. VoIP security technology includes signaling security, media security implements Secure Real Time protocol. Voice communication protection level consider baseline protection level for internal use, advanced protection level for confidentiality, sophisticated protected level for strict confidentiality. In [15] the authors have focused on configuration of firewall to the VoIP network to make it more secure against the attacks coming towards the network. In [16] the authors have concentrated on the security of the VoIP networks with the help of the Virtual Private Network with Internet Protocol Security (IPSec). The idea in [8] [16] has proposed the system architecture that includes three phase. In first phase, the user is registered in phone with the help of sip.conf file. During second phase, the VPN is established by configuring IPsec.conf file so the traffic can be passed through the secure tunneling mode. The last phase consists of the installation of the VPN capable Firewall using IPSec between SIP user agents and switches. The firewalls use Linux as their operating system and open-source firewall software IP Chains and open-source VPN IPSec software FreeS/WAN. In the proposed model of our system the VoIP facility is provided along with IVR (Interactive Voice Response) as a case study that is implemented by developing attendance management system. The VoIP server alone is providing multiple facilities like Mail server with IDS and IPS system implementation, VPN server and firewall iptable rules by configuring it once and it can capable to handle the load of multiple users registered inside the SIP proxy server. Mail server provides multiple facilities to the users of the VoIP system with IVR system by sending them mail in the case of absence. It always updates the system admin about every good and bad request coming towards the VoIP server to use or disrupt its service with the help of OSSEC which is an open source and acting as Intrusion Detection and Intrusion Prevention System. Thus the all in one facility in the VoIP network creates such a precious application which can be desired for any kind of system. The proposed system provides the VoIP facility on Linux Centos 5 platform which is open source operating system. The concept of Virtualization becomes much clearer to the developer by using such a good low cost application.

## 3. PROPOSED WORK

The proposed system architecture is implemented in Virtualization using VMware Tool. In the proposed system architecture the VoIP network is implemented by configuring the files namely sip.conf, extensions.conf and voicemail.conf included in the asterisk 1.6.1.2 package on the Centos 5 Linux platform. Having configured the VoIP network, the VoIP is configured with IVR (Interactive Voice Response) as a case study. On the complete configuration of the mentioned file the system is configured with a mail server which will be responsible for monitoring the asterisk server and alerts the administrator that always monitors and keeps track of the whether asterisk server is being attacked by some blacklisted IP address or not. The system is configured with security parameters in terms of firewalls, Virtual Private Network (VPN), OSSEC (OS Security), Database Security. The phases of the whole system are discussed in a very zoom view. The proposed system includes following phases. Each phase is described in a very zoom view along with the diagrams.

### 3.1. VoIP System Implementation

The whole system is implemented inside the virtual environment of VMware. The system works well for a Local Area Network of an individual environment. Every individual building may configure their own VoIP system for doing inter user calling to their employees.

**(a).Configuring the SIP proxy server**

The implementation starts by configuring the SIP proxy server. For configuration of the SIP proxy server SIP (Session Initiation Protocol) is used. The SIP server can be configured by configuring the sip.conf file path resides inside the /etc/asterisk/sip.conf. This file contains the SIP user registration for Inter Asterisk Communication for VoIP networks. This registration includes various parameters related to user are type, username, host, secret, dtmfmode, insecure, canreinvite, nat, qualify, mailbox, context etc [17]. In the configured sip.conf file these parameters have been taken. Sip file is basically used for the audio streaming. Every user who is using VoIP service inside the Local Area Network (LAN) must be registered within this file. Failing which the user will not able to make inter user calling to other party that is registered in the sip.conf file. In a LAN using a VoIP service using asterisk server will have multiple users obviously. All the users must be registered inside the same context. The mailbox option is configured for sending the user a mail inside its Microsoft outlook express account when the user is not in the state of picking up the call. The mail box will be discussed in the later section. Please note the sip server is using the eth:0 IP address of the Linux system which will work for asterisk server's IP. In the implemented system the private range of IP address has been set which is 192.168.100.37.Asterisk service listens on the default port 5060.

**(b). Making Dial plans**

The Dial Plans are most crucial things inside the VoIP system implementation and they are written inside the extensions.conf [18]. The extensions.conf file path resides inside the /etc/asterisk/extensions.conf. The dial plans guides the asterisk server what to do and how to do. The dial plans are read by the asterisk server very first. Thus the asterisk server is used to read out the extensions.conf file when asterisk is started for implementation. The dial plans are written inside the specific context. Every context has its own different dial plans. When asterisk server reads the dial plans and notices any activity to be done then it first check inside which context the dial plans have been written. On finding the context the asterisk server then reads out the sip.conf file and matches the context written inside the extensions.conf. If both the contexts are getting matched then only the server will execute any specified task without any error. On failure of finding the context inside the sip.conf file, the server will simply return error and terminate the running procedure. The dial plans fill be containing following structure.

[context-name]

exten = > 111 , 1, Operation();

In the above structure [context-name] is the name of the context specified inside the sip.conf. exten is the keyword which is used to write the dial plans inside the extensions.conf. As the file is extensions.conf so every dial plan must be written by using exten = > word. 111 is the extension number of the user to whom the other user wants to call. 1 is the priority number. Using this priority number the task must be performed. The lowest the number of priority the highest the priority is given to the task. Operation() is the function which is the task instructed to the asterisk server which must be done. The Operation() includes various functions of the asterisk like Playback(), Hangup(), Read(), Goto(), Dial(), MYSQL(), SayDigits(), VoiceMailMain() etc. The Interactive Voice Response has been implemented using this file which will be discussed later.

**(c). Configuring voice mail file**

This file is basically used to leave voice message inside the users' x-lite account. When user is dialing specified extensions to it, the voice mail can be listened by him dropped by the caller [19]. The voice mail file also helps the user to leave the voice mail inside its outlook account. This file can be configured by using voicemail.conf file that resides in /etc/asterisk/voicemail.conf. The structure of the voicemail.conf file is as follows.

Inside the extensions.conf file this must be written as follows.

exten = > 444, 1,VoiceMailMain(756@vmail)

Above structure says that 444 is the extension number of X-Lite dialing which the user will be able to listen the voice mail. 1 is the priority number and VoiceMailMain(756@vmail) is the function that must be read out by the asterisk server when it is to be instructed to drop voice mail inside the user's Microsoft Internet outlook express account [19].

Inside the voicemail.conf file this must be written as follows.

756 = > 1234, username, username@domain.com

In the above structure the extension number of the user in the x-Lite is 756. This number must be matched with the extensions.conf. 1234 is the password of the user in the x-Lite to listen the voicemails. Username and user's e-mail with domain needs to be specified. When all these configurations completes the user will be able to listen the voice mail on dialing the X-Lite voice mail extension number. Voice mail provides flexibility to the X-Lite user same as in the Public Service Telephone Network telephone line (PSTN line) in which user can retrieve his/her message on the availability.

**(d) Configuring the X-Lite on the windows client side**

X-Lite is the software which is being used as soft phone in the VoIP implementation. The X-Lite software should be installed on the client side so that every client registered inside the sip.conf on the server can be given their separate extension number with their passwords. The clients on the

223

windows must also be given their voice mailbox password for getting their individual voice mails dropped by different callers. The basic VoIP system is shown in the diagram as shown in figure 1.

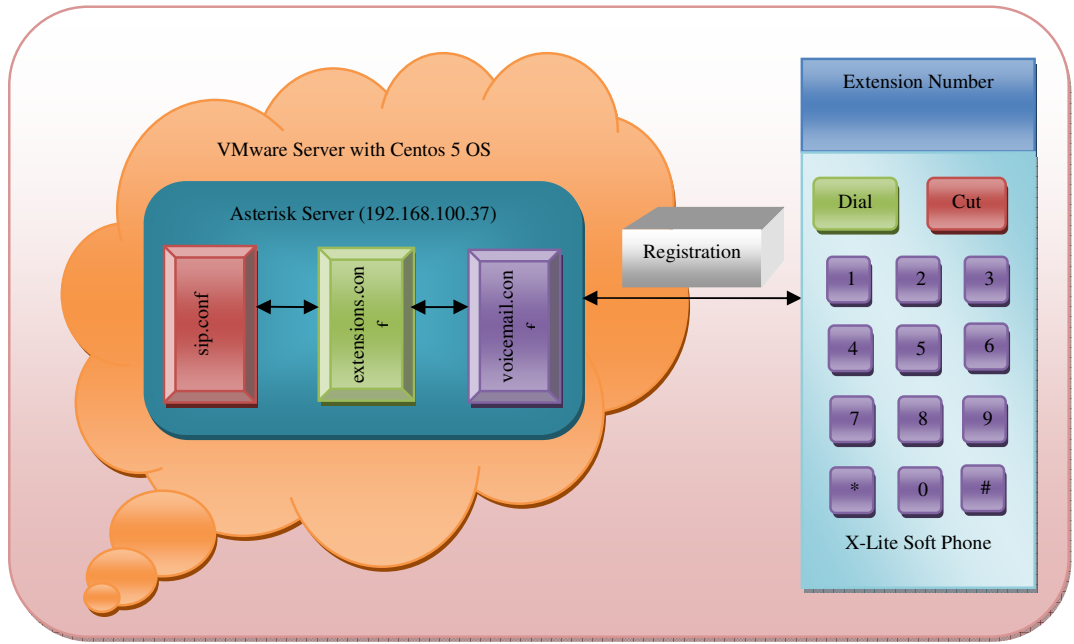

Figure 1.Asterisk Server with sip.conf, extensions.conf, voicemail.conf files connected with X-Lite Soft Phone in a basic VoIP system

As shown in diagram 1 the Asterisk Server (192.168.100.37) is implemented in virtualization inside VMware server environment on Linux Centos 5 platform. Asterisk server will run on Linux platform and X-Lite soft phone is installed on the Windows as client side. Running the asterisk server will help X-Lite to be run and the users on the client side can call each other registered inside the sip.conf.

### 3.2 Implementation of the Interactive Voice Response (IVR)

In the proposed model the Interactive Voice Response (IVR) system has been implemented with the implementation of the VoIP network by installing asterisk add-ons which will be useful to establish connection with MySQL database [26] [28]. Using the IVR system the attendance management system is implemented with MySQL database connection. MySQL database listens on the port 3306, so this port must be opened at the time of the connection. The connection of the asterisk server with the MySQL database is provided in the extensions.conf file while writing the dial plans [27]. With the help of IVR system for any organization the user on the client side those are using the X-Lite soft phone on their respective computers; they will have the advantage of the multiple facilities on the complete configuration of the IVR system. Consider in the deemed university or in the college there must be more than 1500 students. The professors of the university have X-Lite configured on their individual computers and the asterisk server is running by configuring the VoIP system as well as the attendance system with IVR. The professors are getting the intercom facility with the help of the asterisk server running on the Linux platform as well as they can keep track of the final attendance of the student at the time of giving the internal marks by just dialing the specific extension number on the X-Lite phone. The main advantage using IVR system is that the system administrator has to maintain the database records day to day if it is attendance system. In the case of the semester result of the students the database record is updated once during the end of the semester when the result is declared. In the implementation of IVR system I have implemented the student attendance management system using MySQL database connection. My own voice is configured inside the IVR system which will ask the user to enter the ID and password of the student. If the ID and password of the student both are correct then the IVR system will give the response by reading the attendance value stored inside the database. The value is read from the database by only starting the mysqld service which is the necessary service to start the database services. Thus this system can be implemented on any university building. The extensions.conf file also maintains the extensions for using VoIP service as inter user calling also. Here both the functionalities the inter user calling and IVR facility is provided by configuring one single file extensions.conf. The IVR follows the algorithm for implementing the attendance management system as explained in table 1. The asterisk service must be started to start the asterisk server. The mysqld service must be started to establish connection between the asterisk server and MySQL database as well as for fetching the attendance from the database. The postfix service must be started to send the mail as well as the dovecot service is for receiving the mail inside the user's Microsoft outlook express. The description about the postfix and dovecot services will be discussed in the next section on implementation of mail server.

Table 1. The IVR Attendance Management System algorithm with the VoIP inter user calling facility

| Steps | Procedure |
|---|---|
| Step 1 | Start the asterisk service; |
| Step 2 | Start the mysqld service; |
| Step 3 | Start postfix service; |
| Step 4 | Start dovecot service; |
| Step 5 | Play welcome file ; |
| Step 6 | Ask user to enter the student ID; |
| Step 7 | Read student ID entered by the user; |
| Step 8 | Ask user to enter the student password; |
| Step 9 | Read password entered by the user; |
| Step 10 | If(ID= = student ID && password = = student password) |
| Step 11 | Establish the connection between the asterisk server with MySQL attendance database; |
| Step 12 | Search the query based on user demand from the database; |
| Step 13 | Fetch the attendance from database by playing the attendance file; |
| Step 14 | If user still wants to know information about the another student |
| Step 15 | Go to step 6; |
| Step 16 | Else |
| Step 17 | Disconnect the connection between asterisk server with MySQL attendance database; |
| Step 18 | Else |
| Step 19 | Play bad password file; |
| Step 20 | Go to step 6; |
| Step 21 | End If |
| Step 22 | End If |
| Step 23 | Dial extension number of another user |
| Step 24 | If user not replying the call |
| Step 25 | Drop the voice mail to the user |
| Step 26 | Else |
| Step 27 | Complete the call and do Hang up; |

| Step 28 | End If |
| Step 29 | Stop dovecot service; |
| Step 30 | Stop postfix service; |
| Step 31 | Stop mysqld service; |
| Step 32 | Stop asterisk service; |
| Step 33 | End of the algorithm; |

### 3.3. Implementation of Mail Server

Mail server is the crucial portion of the proposed system. With the help of mail server implemented on the asterisk server's IP address (192.168.100.37) the system administrator that is responsible for monitoring the system with the help of security features like firewall, OSSEC, VPN will be notified by the mail to his Microsoft outlook express account. The security parameters used inside the proposed model will be discussed in the next section. The mail server mainly uses two protocols namely Simple Mail Transfer Protocol (SMTP) and Post office Protocol (POP). SMTP listens on the port 25 where as POP listens on the port 110. There are two more protocols which may be used on the implementation of the mail server system namely Secure Simple Mail Transfer Protocol (SSMTP) for sending the mail and POP3S (Post Office Protocol Secure) for receiving the mail. SSMTP mainly listens on the port 225 where as POP3S listens on the port 995. However, in the proposed system the mail server is implemented by using the SMTP and POP protocol. It is very much fruitful to use isolate protocol for sending and receiving the mail individually otherwise the mail server may get loaded heavily if only one protocol is used for sending and receiving the mails. Mail server consists of mainly three things namely Mail User Agent (MUA), Mail Transfer Agent (MTA) and Mail Delivery Agent (MDA) [20]. The MUA is the program which the user uses to read and send e-mail. It reads incoming messages that have been delivered to the user's mailbox, and passes outgoing messages to an MTA for sending. Well known examples for MUA are elm, pine, mutt in UNIX E-mail system. The MTA basically acts as a "mail router". It accepts a message passed to it by either an MUA or another MTA, decides based upon the message header which delivery method it should use, and then passes the message to the appropriate MDA for that delivery method. The well known examples of MTA are qmail, sendmail, postfix, exim [21]. The MDA accepts a piece of mail from an MTA and performs the actual delivery. The main focus behind the implementation of the mail server is to concentrate more on MTAs which are qmail, sendmail and postfix. As qmail is not an open source MTA and its flexibility is good if the study is done on it too hard so it is preferred less in MTA. Whereas in the case of sendmail the administration is a bit tougher, its security is low compared to postfix and it is complex rather flexible. Due to these reasons, the sendmail is rarely used in MTAs. In the proposed mail server the postfix is taken as Mail Transfer Agent as it is IBM's public license, it is used as free open source MTA. It provides very much flexibility if administrator wants to do some crucial changes in MTA. Its security is good as well as administration is easy compared to sendmail. The postfix is used in the proposed model for sending the mail by configuring the file main.cf [22] [23]. This file has a path that is /etc/postfix/main.cf. The main.cf file is configured using hostname, domain name like various parameters. On configuring the file main.cf the file which must be considered for receiving the mails is dovecot.conf. This file has a path /etc/dovecot/dovecot.conf. This file uses the Post Office Protocol which listens on the port 110 so configure this file by keeping port 110 open. On configuring the main.cf and dovecot.conf files the postfix and dovecot service must be started so the mail server can be started on the Linux. On doing this the user has to register his domain and email address on the Microsoft Internet Outlook express. Having done this procedure the connection is established between outlook express on the client side and the mail server inside the Linux successfully. The mail server plays important role to alert the administrator when the security parameters are configured on the proposed model. The complete diagram of the VoIP system with IVR and Mail server is shown in figure 2. The asterisk PBX server sends the voice

226

mail message to the user's Microsoft outlook E-mail address when the call is not picked up with the help of this mail server. As shown in the figure 2 the Asterisk Server having IP address 192.168.100.37 will be working as a Mail server on the same IP. The important point to note the asterisk server is connected with MySQL attendance management system database which provides IVR facility to the X-Lite users along with the inter user calling facility.

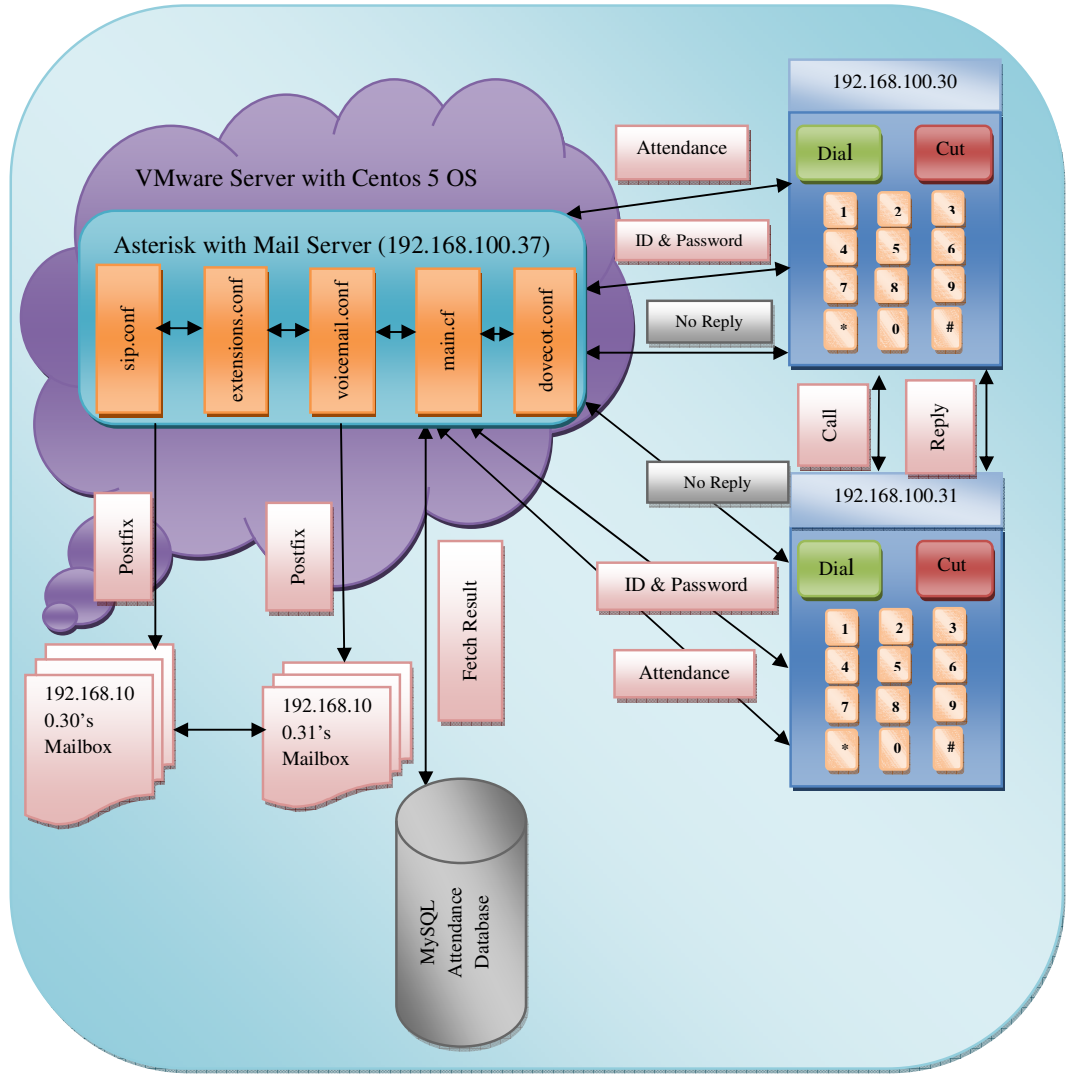

Figure 2. Asterisk Server connection with MySQL Attendance Database, working as Mail Server

The X-Lite users are able to call and communicate with each other by making a call and replying. When any user is not picking up the call, this will be notified to asterisk server which drops the voice mail to the desired user's mailbox account in Microsoft Internet Outlook express. Thus the asterisk server alone is capable to handle the load of IVR server as well as mail server.

### 3.4. Implementing Security parameters

In the proposed system security is required at these levels namely Asterisk Server Security, Mail Server Security, Database Server Security. As Asterisk server behaves as a mail server also, it must be secured by configuring the security parameters to it. This can be achieved by implementing the firewall, VPN server, Intrusion Detection and Prevention system (IDS & IPS). Each parameter is discussed in a precise view.

### 3.4.1. Asterisk Level Security

Asterisk server except providing the VoIP and IVR facility to clients, works as a Mail Server. The security to asterisk server must be very much necessary. The security can be provided to the asterisk by configuring it with a Virtual Private Network (VPN) server.

### (a). VPN Server configuration

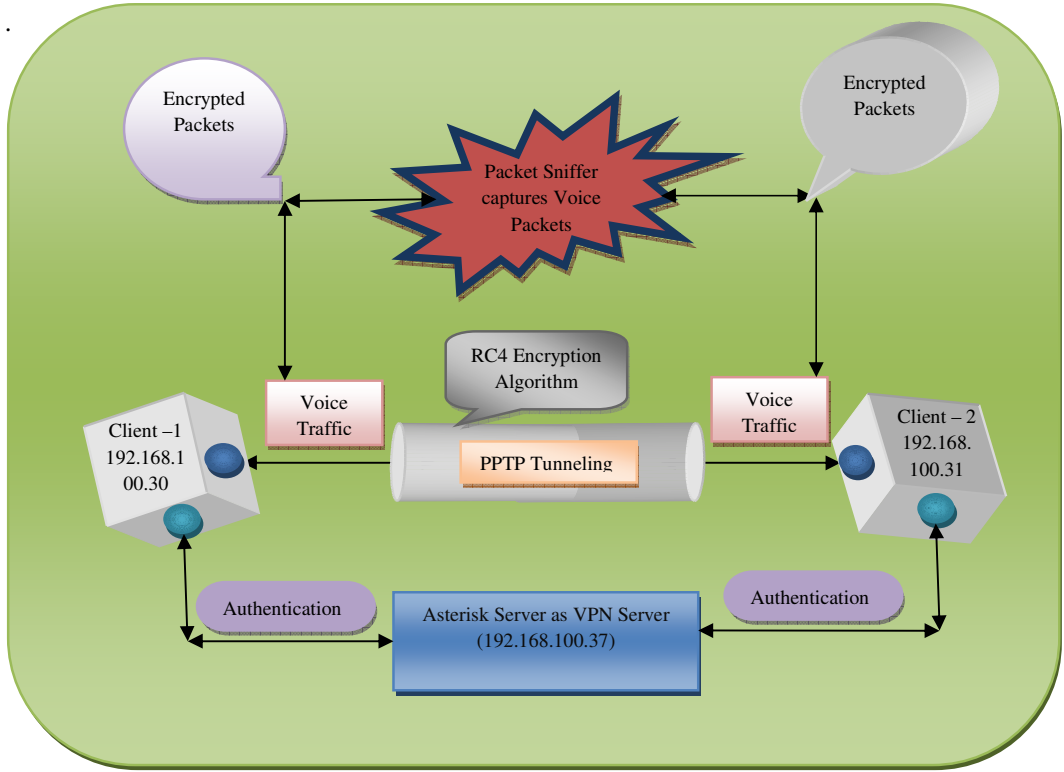

Figure 3.Asterisk Server as VPN server uses PPTP tunneling for Encrypted Voice packets

Virtual Private Network (VPN) is the network that provides the secure access to the remote offices or travelling users to a centrally deployed network. VPNs require remote users of the network to be authenticated and often secure the data with encryption technology to avoid the unnecessary disclosing the private information of the organization to the third parties. Virtual Private Network typically uses various tunneling protocols namely Point to Point Tunneling Protocol (PPTP), Layer 2 Tunneling Protocol (L2TP), Secure Socket Tunneling Protocol (SSTP) for Point to Point Protocol [37]. In the proposed system PPTP is used for sending the VoIP traffic from one client to the other client and traffic from X-Lite users to asterisk server is passed to the tunnel created by the PPTP. The exact scenario is discussed here. Initially the VPN is configured using the pptpd.conf. This file has path /etc/pptpd.conf. Using this file the local IP and the pool of remote IP range is set. Outside the range of remote IP the VPN will not allow any other IP to be registered inside it. On the complete configuration of pptpd.conf file the password of the user is set by configuring the file chap-secrets to be registered inside VPN. The chap-secrets file has the path at /etc/ppp/chap-secrets. With the help of chap-secrets the passwords are stored for the specific list of users who are allowed to be registered inside the established Virtual Private Network. The VPN service is started by starting pptpd service. The VPN server is providing the eth:0 IP of the asterisk server which was set to 192.168.100.37. Every client (X-Lite user) must be registered with VPN server by providing their IDs and Passwords which have been set to chap-secrets file. When the pptpd service is started, after that the asterisk service must be started so every SIP user can be registered to the X-Lite. The advantage for doing so is the voice traffic will be passed through the PPTP's tunnel [24] [25]. PPTP tunneling uses RC4 encryption algorithm with 128-bit key. Every voice packet will be travelled through the secure tunnel in the encrypted form. Even if the packets of the voice are captured by the packet sniffer the packets will be in encrypted form. In this configuration the asterisk server behaves as VoIP service server, Mail Server as well as VPN server for securing the data to be sent securely inside the tunnel from one end to another end. Virtual Private Network server configuration as Asterisk Server is shown in figure 3. The clients of the VoIP systems are X-Lite users.

**(b). Firewall configuration**

The firewall is a device or set of devices which allows or deny requests coming to the server based on the some set of rules to defend the network against any kind of unauthorized access. Firewall basically works on the IP table rules. With the help of the rules the system administrator can configure the firewall to protect the server so network will be protected automatically. According to the IP table rules the packets must be either in ACCEPT, DROP or REJECT state. The firewall is implemented for protecting the asterisk server against malicious request coming towards it. As discussed in the previous sections the asterisk server, the MySQL database server, SMTP, POP, the VPN server listens on 5060, 3306, 25, 110, 1723 ports respectively. For using the VoIP & IVR system with Mail server and VPN server these ports must be kept open. As VoIP is Voice over Internet Protocol, the main cup of tea in the application is audio streaming. The users on the VoIP service can interact with each other using voice as a communication medium. In the case of audio, video streaming the User Datagram Protocol (UDP) is involved. Due to this protocol, the ports related to UDP also must be kept open. Remaining ports related to protocol like TCP must be blocked. The icmp port is kept open to check whether the clients are connected with the asterisk server or not. By keeping SSH port which is 22 open, the remote login can be done by the administrator to monitor the asterisk server from any computer in the LAN of an organization. To configure the firewall following rules must be fired to defend the network against unauthorized access is shown in table 2 [35] [36].

Table 2. IP table Rules for configuration of Firewall to defend asterisk server

| Protocol | iptable rules |
| --- | --- |
| TCP | iptables –I INPUT –p tcp –j DROP |
| SSH | iptables –I INPUT –p tcp –dport 22 –j ACCEPT |
| ICMP | iptables -I INPUT -p icmp -j ACCEPT |
| UDP | iptables -I INPUT -p udp -j DROP |
| SIP | iptables -I INPUT -p udp --dport 5060 -j ACCEPT |
| MySQL | iptables -I INPUT -p udp --dport 3306 -j ACCEPT |
| PPTP | iptables -I INPUT -p tcp --dport 1723 -j ACCEPT |
| PPTP | iptables -I INPUT -p udp --dport 1723 -j ACCEPT |
| SMTP | iptables -I INPUT -p tcp --dport 25 -j ACCEPT |
| SMTP | iptables -I INPUT -p udp --dport 25 -j ACCEPT |
| POP | iptables -I INPUT -p tcp --dport 110 -j ACCEPT |
| POP | iptables -I INPUT -p udp --dport 110 -j ACCEPT |

**3.4.2. OSSEC Security with the Mail Server**

As asterisk server is configured to work as a Mail server also the security to this module is also mandatory thing in this proposed system. The system administrator of the network continuous monitors the network with the help of the mail server hence any malicious which are not blocked by firewall must be notified to the administrator so the necessary actions can be taken in no time.

**(a). VPN Server configuration**

To provide more security in VoIP system the mail server is configured with open source tool which is OSSEC (OS Security). OSSEC is configured in a way that will continuous observe the request coming from while list of IP range which are registered inside VPN. If more than certain specified requests say 10 requests with in specific time quantum come to asterisk server the OSSEC will black list that IP address and notify to the system administrator immediately about the blacklisted IP address. Along with the notification sent to the admin, the OSSEC will block that blacklisted IP address of the client as an immediate response from the server [32]. This configuration has to be set inside the OSSEC by configuring the firewall script inside the directory /var/ossec/active-response/bin/firewall-drop.sh. OSSEC is client-server architecture compatible tool for serving itself as IDS and IPS system. With the help of OSSEC tool the asterisk server can be protected against Distributed Denial of Service (DDoS) attack in a very efficient manner. This tool will behave like Intrusion Detection System (IDS) and Intrusion Prevention System (IPS) [30] [31]. Mail server is configured in such a way that without the authentication of the specific user it would be a bit tougher for the attacker to access the user's Microsoft outlook mail express account. Thus with the help of the Mail server and OSSEC tool the administrator of the system will be getting mail from the OSSEC about the blacklisted IP addresses, also from asterisk PBX about the voice mail dropped to callee by a caller. The OSSEC generates different security alert levels ranging from 0-15 which are explained in table 3. These levels should be read from lowest level to highest level [29].

Table 3. Various alert levels provided by OSSEC

| Alert Level | Action Taken |
|---|---|
| 00 | It should be ignored hence no action should be taken. It is used to avoid false positive. |
| 01 | No action should be taken on generation of this alert level. |
| 02 | It is used to generate the system notification or status messages. This level has less security relevance. |
| 03 | It is used to monitor successful events. It includes firewall allow packets, successful login attempts, etc. |
| 04 | It is used to indicate that the system low priority error. This level includes the errors related to bad configurations. |
| 05 | It is used to pretend user generated errors. This includes missed passwords, denied actions. They have no security relevance. |
| 06 | It is used to expose low relevance of attack. They indicate a worm or a virus that have no affect to the system (like code red for apache servers, etc). They also include frequently IDS events and frequently errors. |
| 07 | It is used to indicate "Bad Word" matching. This alert level has no security relevance. |
| 08 | It is used to describe first time seen events. This includes packet sniffing kind of activity. IDS event fire when first time observed event occurs. |
| 09 | It is used to indicate the error from the invalid source. This includes the login as unknown user or from invalid source. It includes the errors related to admin (root) account. |
| 10 | It is used to show the multiple user generated errors. It typically includes multiple bad passwords, bad logins. This level indicates the errors are attack or users have forgotten their credentials. |

| 11 | It is used to describe the integrity check warning. It includes the messages regarding the modification of binaries or the presence of root kits (by root check) |
|---|---|
| 12 | It is used to expose highly importance event. They include error or warning messages from the system, kernel, etc. They may indicate an attack against a specific application. |
| 13 | It also used to indicate the unusual error with high importance event. Most of time it matches common attack pattern. |
| 14 | It is a high importance security event, and it indicates an attack by making co-relation with time. |
| 15 | It is used to describe for severe attack on the server and no chance for false positive and immediate actions must be taken to prevent the attack. |

On getting the above mentioned alert levels the administrator takes necessary actions based on the priority levels of alerts. The alert levels from 8-15 indicate the severity of attacks and actions to be taken immediately where as the alert levels from 0-7 has lower priority in terms of actions to be taken by the system administrator. OSSEC works well with the agents like VMware, HP-UX, Solaris, Microsoft and Linux.

### 3.4.3. Database Level Security

In the proposed system as the database which is in connection with asterisk PBX server is MySQL, the security vulnerabilities are still possible to the database also. If the attack is done to the database, it may lead failure of the IVR service in the VoIP network. The private information about the users may get lost about their ID and passwords etc. The database must be made secure with the help of the proper authentication procedure. The fully read and write privilege should be allowed to top level management of an organization with system administrator of the network. Remaining any lower level management employees must have only read privileges to the tupples of the database. Doing so the lower level employee of the organization can't make any updates in the database. The system administrator will have its administrative privileges and database access privileges. The administrator contains the administrative privileges like CREATE TEMPORARY TABLES, FILE, GRANT OPTION, LOCK TABLES, PROCESS, RELOAD, SUPER, SHUT DOWN, SHOW DATABASES, REPLICATION CLIENT, and REPLICATION SLAVE. He will have the database access privileges which he will be providing to the users of the systems are namely ALTER, CREATE, DELETE, DROP, INDEX, INSERT, SELECT, UPDATE. The admin provides the privileges to the specific user at specific IP using GRANT command and cancels the privileges to the blacklisted user by executing REVOKE command. Thus providing the GRANT and REVOKE privileges to the users for accessing the database the security can be enhanced in a very fruitful manner. The example of providing the specific privilege to specific user is shown below [33] [34].

$ mysql –u abc –pabc;

The above command fired on the mysql prompt will create the specific user named abc and the password set for the user abc will be abc by the system administrator. It is shown in figure 4.

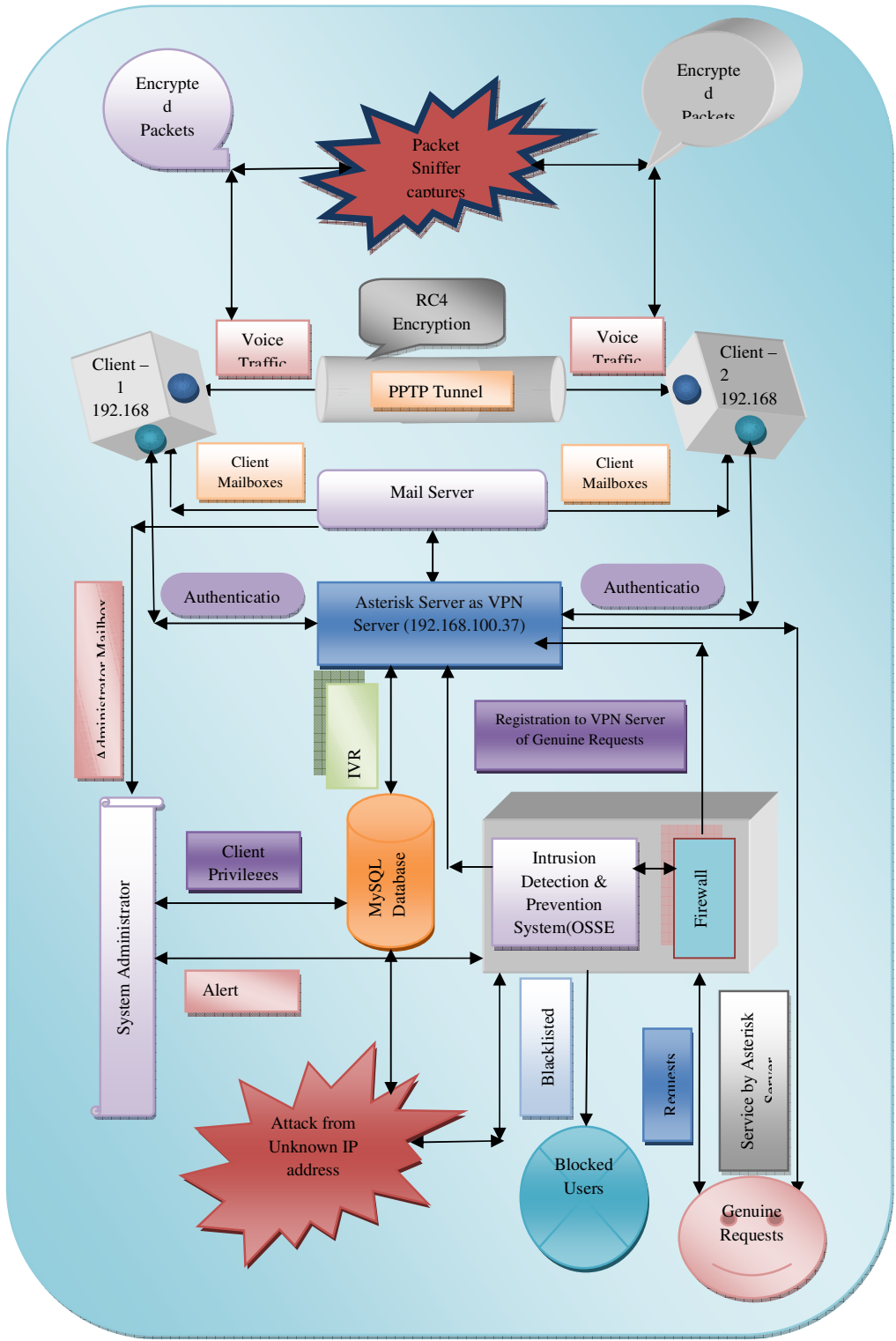

Figure 4. Complete Proposed System architecture with security parameters attached to the Asterisk Server in VMware server environment on Linux Centos 5 platform.

232

International Journal of Distributed and Parallel Systems (IJDPS) Vol.3, No.3, May 2012

$ GRANT INSERT,UPDATE, ALTER on p.q TO 'abc'@IP IDENTIFIED BY 'abc';

On giving the above command the system administrator will provide INSERT, UPDATE and ALTER privilege to the user abc. Identified by 'abc' indicates the password for user abc is abc. p is the name of the database to which abc will get access and q is the table belongs to the database p. IP indicates the user is residing on the specific IP. Similarly all privileges can be provided to the user abc is written by the command given below.

$ GRANT ALL on *.* TO 'abc'@IP IDENTIFIED BY 'abc';

*.* indicates all the tables in all the databases will be accessible by the user abc with password abc at specific IP. ALL indicated the user abc will have all privileges to all the databases with all tables. Similarly the privileges can be taken back to specific users by giving REVOKE command.

$ REVOKE ALL on *.* TO 'abc'@IP IDENTIFIED BY 'abc';

$ REVOKE DELETE, DROP, ALTER on x.y to 'abc'@IP IDENTIFIED BY 'abc';

GRANT and REVOKE command helps the system administrator to provide the specific privileges to the specific user in a flexible manner. The complete proposed architecture with firewall, mail server with OSSEC and database security implementation with asterisk server & IVR system is shown in figure 4. In the diagram the firewall and the OSSEC are placed at same level. Firewall accepts drops or rejects the packets if they are coming from the unauthorized users while OSSEC monitors the events and if necessary the firewall scripts may be configured to OSSEC by the system admin as immediate actions taken by OSSEC before system administrator takes. Configuration of firewall and OSSEC will help to administer the system in the physical absence of the system admin. The connection between OSSEC and mail server is also shown in the diagram 4. As explained in figure 4 the genuine requests are passed through the IDS and IPS system attached with firewall and registered with the VPN server which is asterisk server itself. On completion of authentication the VOIP service is provided to the requested user. If attack is done to the asterisk server the requests of the attacker must be passed from the IDS and IPS system attached with firewall and based on the configuration set in the system by the system administrator the bad request will be denied by the server and list of the blocked users is maintained with IP addresses so in the future if any request comes from the blocked user list, the service can be denied immediately. The asterisk server is connected with IVR system, Mail Server, IDS & IPS system with firewall system and configured itself as a VPN server.

## 4. PROS AND CONS

The main advantage of the proposed system is the asterisk server alone behaves as VOIP server, VPN server for tunneling the traffic using Peer to Peer Tunneling Protocol (PPTP), Mail Server to alert the system administrator if any attack is going on as well as to alert the X-Lite users in their absence. Asterisk server is capable enough to provide the IVR service to the window clients by establishing the connection with MySQL database very efficiently. The Mail server configured by the asterisk server helps the system administrator a lot in configuring OSSEC and Iptable rules very efficiently. The multiple servers' configuration facility with asterisk server along with the security parameters makes the application very much well-organized for a large organization using VOIP service. Any organization can develop the attendance management system; the result management system can be created by the college or deemed university, the train arrival timing application based on asterisk by the railway department very easily. However the attendance management system has been implemented for the university system by me in the proposed work.

233

International Journal of Distributed and Parallel Systems (IJDPS) Vol.3, No.3, May 2012The development of the VOIP and IVR system has been implemented on the Linux Centos 5 platform using asterisk and MySQL packages. The important point to note is the asterisk and MySQL packages are open source packages. These packages can be available to any user very easily. In fact the Centos 5 is also an open source operating system. The X- Lite can be purchased by paying very economical money. VPN server configuration is created by PPTP configuration as discussed earlier which is also an open source configuration file. Similarly the OSSEC is an open source tool for configuring intrusion prevention and detection system. The Linux system inbuilt provides the iptable rules so firewall can be configured by firing the rules of iptable. The postfix and dovecot configuration files used for configuration of the mail server can be easily available as they are also easily available due to open source files. Thus any organization can develop the VOIP application by just configuring the necessary files mentioned in the proposed system in very proficient manner. The open source packages like asterisk, MySQL, PPTP, OSSEC, Firewall iptable rules provide a cost effective VOIP solution to the organization.

The proposed system architecture removes the need for having hard core telephone landline instruments to be used as intercom for communication between two parties by providing X-Lite based IP soft phones so this provides space efficient as well as wire efficient solution to the organization as well as their employees.

The proposed architecture helps system admin to protect against brute force attack, Distributed Denial of Service attack (DDoS) on asterisk server, SIP registration hacking attack and database attacks on MySQL database, port scanning attacks and Registration Hijacking attacks in very fruitful manner.

The main limitation of the asterisk system is that it only deals and understands the number. It doesn't understand the string or set of letters provided as the input by the users. Hence in any system the input provided to the asterisk system must not be string as ID or as Password in any case. This limitation can be avoided by taking or providing the inputs in the form of pure numbers.

## 5. SNAP SHOTS OF PRACTICALS PERFORMED

Snap Shot 1. Asterisk server provides communication service using SIP and RTP protocol using Wireshark packet capturing tool it is shown clearly.

International Journal of Distributed and Parallel Systems (IJDPS) Vol.3, No.3, May 2012

Snap Shot 2 When client is registered to VPN immediate compressed data is started to passes

Snap shot 3. Now the client is registered with new IP of 192.169.100.10 instead of 192.168.100.36. This is the power of VPN server. The asterisk server is now connected with new IP of client provided by VPN server.

## 6. CONCLUSION

The proposed system architecture has been implemented in the VMware Server's virtualization environment on Linux Centos 5 platform using Bridge networking for a Local Area Network (LAN) in a university Lab. The asterisk package is very much easily available and configurable on a Linux platform. The proposed system provides a cost effective VOIP and IVR solution to the organization having thousands of employees. The proposed architecture has been implemented for whole university environment. Similarly if any organization is having multiple departments and if it wants to develop the VOIP service for individual department, in such a scenario two asterisk servers must be registered to each other by configuring iax.conf file that can be developed as a future aspect. In the proposed architecture the phone which are being used for inter user communication are X-Lite based IP soft phones. This system can be implemented on the hard core telephone instruments using digium telephony card which must be inserted in to CPU port as a future aspects. The proposed architecture removes the need of having hard core telephone instruments those must be there for inter com kind of communication by providing soft phone communication as intercom. The proposed architecture just requires the head phones for communication between computer to computer if there are the desktops inside the organization connected inside the Local Area Network. If the organization has set up the laptops to their employees, the need for the head phone can even be minimized as the laptops are having the inbuilt mike and speaker which are essential for communication between two computers same as gtalk, team viewer, Skype voice chat. Using the proposed architecture any person can configure his house with an intercom facility in different rooms by just installing X-Lite phones to his multiple computers and configuring the asterisk on one of the computer as VOIP server if the person is having multiple computers with him. Thus, the system architecture represents reveals the magical power of Linux as a open source growing technology.

**Authors**

**Kinjal Shah** received his B.E degree in Computer Engineering from A. D. Patel Institute of Technology, New Vallabh Vidyanagar, Distt. Anand, Gujarat in 2009. He is currently pursuing MTECH degree in Computer Science Engineering and Information Technology at Jaypee University of Information Technology, Waknaghat, Distt. Solan- 173234. He is doing thesis work on security issues in VoIP networks in Virtualization with IVR. His research interest includes Cloud Computing, Cryptography and Network Security and Computer Networks.

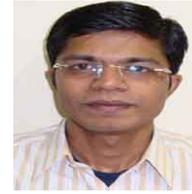

**Prof. Satya Prakash Ghrera** is currently Head of the department of Computer Science Engineering and Information Technology at Jaypee University of Information Technology, Waknaghat, Distt. Solan-173234. He received his B.Sc Engineering (Hons) from Regional Engineering College (REC) now NIT Kurukshetra during April 1971. He served himself in Corps of Electronics and Mechanical Engineers of the Indian Army till 34 years since 1971. He received his ME (Hons) Computer Science from Thapar Institute of Engineering and Technology, Patiala during 1995. He received his MBA from HR and IR department from University of Madras during 2004. He was awarded Army Commander's Commendation twice in 1988 and in 2004 for his distinguished service. Currently he is pursuing Ph.D degree in Computer Science and Engineering. His research area includes Design of Computer networks, Computer and Network Security, Integration of Computer Networks and Communication Systems, Network programming and Management of Network based Real Time Information Systems.

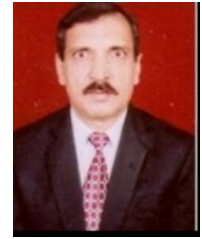

**Alok Thaker** is currently working as Linux/Network/VoIP/Security Consultant at Inferno Solutions, Vadodara, Gujarat. He has 6 plus years of IT experience in Linux system development, administration, networking & security being MCA (Masters of Computer Applications) from Sardar Patel University, Vallabh Vidyanagar, Distt. Anand (Gujarat) in 2005. He has been consultant to various firms in India and abroad with onsite visits & data center deployments in Singapore, Newyork, Japan etc. His technical proficiency includes embedded Linux, Production servers on Linux, training, consultancy, VOIP, UTM's (Unified Threat Management Systems), Network security etc. He is also the moderator of largest Linux user group in India called VGLUG.

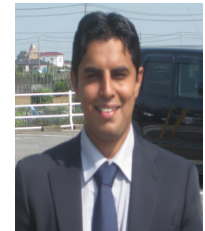